\documentclass{amsart}
\usepackage{amsmath}%
\usepackage{amsfonts}%
\usepackage{amssymb}%
\usepackage{graphicx}
\newtheorem{theorem}{Theorem}
\theoremstyle{plain}

\newtheorem{condition}{Condition}

\newtheorem{corollary}{Corollary}

\newtheorem{definition}{Definition}

\newtheorem{remark}{Remark}

\numberwithin{equation}{section}
\begin{document}
\title[$P=NP?$]{On the principal impossibility to prove $P=NP$}
\author{Natalia L. Malinina}
\address[Moscow Aviation Institute (National Research University), Department of Computing Mathematics and Programming, Russia]
\newline%
\email{malinina806@gmail.com}%
\thanks{This paper is somewhat an extended version of the report, which was made on the International Congress of Mathematicians in Hyderabad, India, in 2010.}  
\date{October 30, 2012}

\dedicatory{Dedicated to my father Leonid Malinin}

\begin{abstract}
The material of the article is devoted to the most complicated and interesting problem --- a problem of $P=NP?$. This research was presented to mathematical community in Hyderabad during International Congress of Mathematicians \cite {Malinina1}. But there it was published in a very brief form, so this article is an attempt to give those, who are interested in the problem, my reasoning on the theme. It is not a proof in full, because it is very difficult to prove something, which is not provable, but it seems that these reasoning will help us to understand the problem of the combinatorial explosion more deeply and to realize in full all the problems to which we are going because of the combinatorial explosion. Maybe we will realize that the combinatorial explosion is somehow a law, such a law, which influences the World, as Newton law of gravitation influences the fall of each thing.
\end{abstract}
\maketitle

\section{Introduction}

The years roll by and our world becomes more and more complicated. Nobody can recall the combinatorial explosion, which was opened by mathematicians approximately half of the century ago. For example, the proof of the Pythagor's theorem takes several lines; the proofs of the present-day theorems may take several pages and sometimes much more. 

A problem to prove or to disprove the equality between $P$ and $NP$ classes of the tasks was raised long ago \cite {Cook} and nowadays it still does not have a strict solution. Let us try to make an attempt to explain the problem on the base of the proved theorems on the solution of the graph isomorphism problem \cite {Malinina},\cite {Malinin}. 

A problem of searching for the duality between the vertex graphs exists in the graph theory and it is named as "Problem of Graph Isomorphism". It was broadly lightened in the book of Gary and Jonson "Computers and Intractability" \cite {Gary and John}.

A concept, more known as the Church-Turing thesis, on the normalization of the algorithms exists in the computer science \cite {Turing}. The theory of algorithms considers that all algorithms are normalized \cite {Markov}. It is sufficient to add only one letter to the alphabet in order to receive a normalized algorithm, but yet we have not got a regular method for the addition of such letters or for the normalization. 

\section{An equivalent duality between graphs}

Further, it is well known that the graph, surely, the directed graph, may serve as the pictorial representation of the algorithm. It is also a well known fact that many problems are solved easily if they can be presented with the help of the edge graphs. We can easily compare the edge graphs, but there are no exact methods for the comparing of the vertex graphs, only the approximate ones.

\begin{figure}[htb]
	\includegraphics[width=0.45\textwidth]{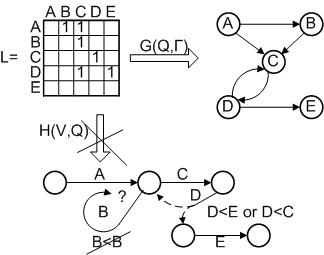}
	\caption{}
	\label{Fig:1}
\end{figure}

The problem on the duality between the vertex graphs and the edge graphs exists in the graph theory. It is known that every edge graph, which is specified with the help of an arbitrary adjacency matrix, can be easily transformed into the vertex graph, but the opposite duality (transformation) appears only occasionally (fig.\ref{Fig:1}). 

So, the duality between the directed vertex and edge graphs was examined. It was interesting to know at what conditions an arbitrary adjacency matrix, will be at the same time the adjacency matrix of the vertexes of the vertex graph and the adjacency matrix of the edges of the edge graph. The answer was obtained in the form of the proved theorem "On the quasi-canonical adjacency matrix" \cite {Malinina1}. The theorem was proved fourty years ago, but according to very specific reasons was published only in 2009 after the death of its author. Let me present its formulation.

A theorem on the quasi-canonical adjacency matrix determines both necessary and sufficient conditions that the direct path's $L$ matrix has a dual nature that is at the same time it might be both $E$ --- adjacency matrix of the $G$ graph's vertexes and $R$ --- adjacency matrix of the $H$ graph's edges on condition that they may have not equal cyclomatic numbers. The theorem was proved for the case of the directed graphs, since any undirected graph $G$ may be transformed into a directed one by the operation of the redoubling \cite {Berge}.

Below you will see the formulation of the theorem. 

\begin {theorem}[On a quasi-canonical adjacency matrix]

It is given: a set $Q=\{q_i\}$ and $L=Q \times Q$ by the way of $q_i<q_j$, and $L=\left\| l_{ij}\right\|_1^n=\left\|e_{ij}\right\|_1^n$ for graph $G(Q,\Gamma)$. Then $\left\|l_{ij}\right\|_1^n=\left\|r_{ij}\right\|_1^n=R$ for graph $H(Q,V)$ if and only if:
\begin {enumerate}
\item
An $L=\left\|l_{ij}\right\|_1^n$  generates $C_n=\left\|c_{ij}\right\|_1^n=[0]$
\item
A minor $|l_{ij}|_1^{(n-1)}$  of every $l_{ij}=1$ generates $C_{(n-1)}=\left\|c_{ij}\right\|_1^{(n-1)}=[0]$
\end{enumerate}

Where:

$c_{ij}=l_{ij}(\delta_{j/i}*s_{ij}+\delta_{i/j}* s_{ij})$; 

$\delta_{j/i}*s_{ij}=(s_{ij}-min_j s_{ij})_i$;

$\delta_{i/j}*s_{ij}=(s_{ij}-min_i s_{ij})_j$;

$s_{ij}=l_{ij}*(\sum_{\stackrel{i=1}{j}}^k l_{ij}+\sum_{\stackrel{j=1}{i}}^k l_{ij})$;

$(min_j s_{ij})_i=min_{j/i} s_{ij} \in \{s_{ij} \neq 0 \}$;

$(min_i s_{ij})_j=min_{i/j} s_{ij} \in {s_{ij} \neq 0}$;

$k=n,(n-1)$;

\end {theorem}

The proof of the theorem takes more than 15 pages \cite {Malinina}, \cite {Malinin}. Both graphs must be the restricted ones, and also they must not have the contours. The problem of the contours will be discussed a bit later. 

Due to the requirements of the structural similarity of the graphs, the following topological invariants must be equal: 

\begin {enumerate}
\item
The numbers of the elements of both the initial and the fundumental sets;
\item
The binary relation's systems, assigned on the fundumentsl sets of the elements;
\item
The cyclomatic numbers (but not in all cases).
\end {enumerate}

\begin{figure}[htb]
	\includegraphics[width=0.95\textwidth]{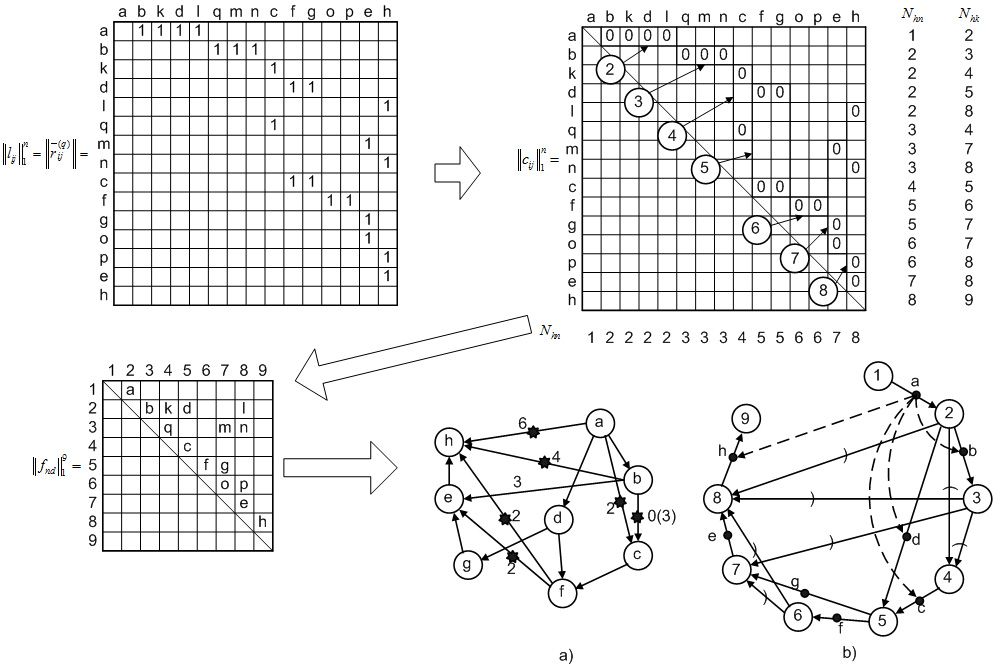}
	\caption{}
	\label{Fig:2}
\end{figure}

The example of the matrix which meets the requirements of theorem 1 is presented in fig. \ref{Fig:2}.

We can call such vertex graph as the quasi-canonical or semi-canonical vertex graph and the edge graph as the quasi-canonical edge graph. The cyclomatic numbers here are not equal, and the duality is not the strict one. 

Another problem arises: how to reduce an arbitrary adjacency matrix to the form, which possesses the needed properties? That's why such a transformation of the adjacency matrix, which does not disturb the transitivity property of the binary relation, was worked in. 

And the theorem "On a quasi-normalization of the $L$ matrix's binary relations" was proved \cite {Malinina}, \cite {Malinin}.

But before let us explain the concept of the $\Delta n$--transformation.

\section{The $\Delta n$--transformation}

Let's change the relation $q_i<q_j$ on two relations: $q_i<q_x$ and $q_x<q_j$. We'll find that $q_i<q_x<q_j$. It is obvious that two relations both $q_i<q_x$ and $q_i<q_x<q_j$ are equivalent according to the initial pair of the elements.
Therefore, the inserting of $q_x$ element into $q_i<q_j$ relation --- is the concervative operation regarding to this relation in the initial pair.
So, we will add one row (both line and column) to $L$ matrix at the condition of replacement $q_x<q_y$ relation with the pair of binary both $q_x<q_{n+1}$ and $q_{n+1}<q_y$ relations.

\begin {theorem}
[On a quasinormalization of the $L$ matrix's binary relations]

Any direct path's $\left\|l_{ij}\right\|_1^n$ matrix can be transformed to the quasi-canonical (quasi-normal) $\left\|l_{ij}\right\|_1^{(n+s_q)}$ form by means of applying the $\Delta n$--transformation to such $s_q$ ($s_q \neq n^2-1$) elements of the $\left\|l_{ij}\right\|_1^n$ matrix, which do not satisfy to the conditions of theorem 1. 
\end{theorem}

This transformation has a very simple graphical interpretation on the initial graph. All the edges, for which $c_{ij}\neq 0$, are divided into two consecutive ones, and the new additional vertexes are inserted into the gaps. 

Let us settle that by the $\Delta n$--transformation of matrix $L$ we will comprehend the addition of one row (both line and column) to matrix $L$ at the condition of replacement relation $q_x<q_y$ with the pair of binary relations both $q_x<q_{n+1}$ and $q_{n+1}<q_y$ \cite{Malinin}, \cite{Malinina}. 

It is evident, that the $\Delta n$--transformation is the conservative operation regarding the binary relation in the initial pair $(q_i;q_j)$ and does not break such a structural similarity criterion as the binary relation's system.

After this operation for all the edges, which are adjacent to the divided ones, the named value is calculated once more, and again the edges, which possess such property are divided into two consecutive ones. The process continues up to the moment, when all such values are equal to zero for all the edges. It was proved that such an operation appears to be the convergent one.

$c_{ij}$ --- is such a quantitie's property of $L$ matrix. Formulas for its calculatinf are presented above.

\begin{figure}[htb]
	\includegraphics[width=1\textwidth]{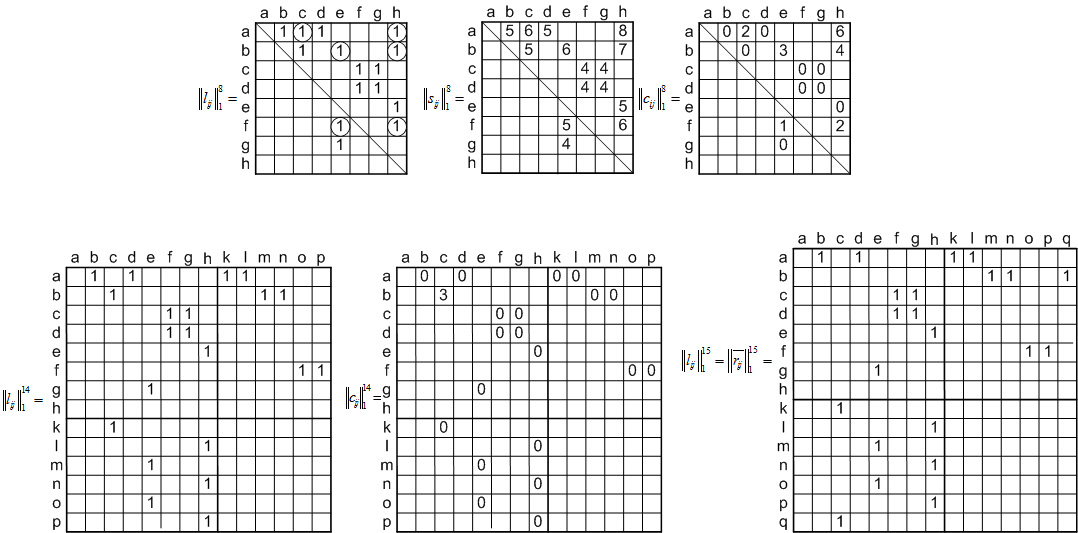}
	\caption{}
	\label{Fig:3}
\end{figure}

\begin{figure}[htb]
	\includegraphics[width=0.55\textwidth]{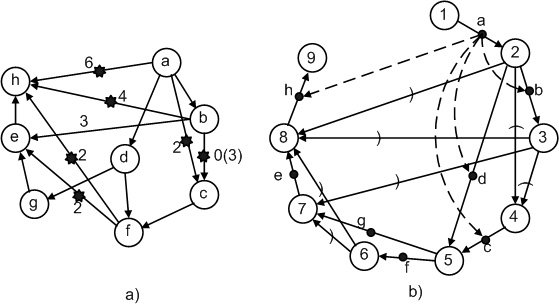}
	\caption{}
	\label{Fig:4}
\end{figure}

The demonstration of the application of the $\Delta n$--transformation to the arbitrary matrix is presented in fig. \ref{Fig:3} and fig. \ref{Fig:4}.

Now let us examine the case of the strict duality of both vertex and edge graphs, when the demand of the cyclomatic number's equality is applied together with other structural similarity criteria, because the cyclomatic number is rather a quantitative estimation of the binary relation's system $L=Q \times Q$. Thus, we have theorem "On the canonical adjacency matrix". Let's bring in some definitions:

\begin{definition}
By graph $H$ simple vertex $v_h$ we'll denote such vertex, which corresponds to submatrix $|r_{ij}|_k^p$ of matrix $R$ at both $k \geq 1$ and $p=1$ or at both $k=1$ and $p \geq 1$.
\end{definition}

\begin{definition}
By graph $H$ elementary vertex $v_h$ we'll denote such a vertex which corresponds to submatrix $|r_{ij}|_k^p$ of matrix $R$ at both $k=1$ and $p=1$.
\end{definition}

\begin{definition}
By graph $H$ complicated vertex $v_h$ we'll denote such vertex $v_h$ which correspond to submatrix  $|r_{ij}|_k^p$ of matrix $R$ at both $k \geq 2$ and $p\geq 2$.
\end{definition}

The full and correct formulation of the theorem you can see below.

\begin{figure}[htb]
	\includegraphics[width=0.45\textwidth]{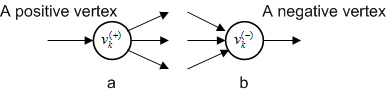}
	\caption{}
	\label{Fig:20}
\end{figure}

\begin {theorem}
[On the canonical adjacency matrix]

Let the initial graph $G$ be specified as matrix $\left\|e_{ij}\right\|_1^n$ --- an adjacency matrix of vertexes, which corresponds to the quasi-canonical matrix $\left\|r_{ij}\right|_1^{(n+s_q)}$ --- an adjacency matrix of edges of the connected edge graph $H_q$. In order that graph $H_q$ cyclomatic number $\nu(H_q)$ might be equal to the initial graph $G$ cyclomatic number $\nu(G)$, it is necessary and sufficient for all graph $H_q$ vertexes to be simple (fig. \ref{Fig:20}, or, for every $r_xy=1$ such condition must be fulfilled:

\begin{condition}
$
\left\{
        \begin{array}{l}
$
                if $\sum_{\stackrel{i=1}{j=y}}^{(n+s_q)} r_{ij}\geq 1$,then $\sum_{\stackrel{j=1}{i=x}}^{(n+s_q)} r_{ij}=1$;
$
\\
$
                if $\sum_{\stackrel{j=1}{i=x}}^{(n+s_q)} r_{ij}\geq 1$,then $\sum_{\stackrel{i=1}{j=y}}^{(n+s_q)} r_{ij}=1$:
                 $\ 
        \end{array}
\right.
$
\end{condition}
\end{theorem}

It may be produced in a very simple form:

If both graphs: $G$ --- vertex, and $H_q$ --- edge, have one and the same initial binary relation system $L=Q \times Q$, adjusted on set $Q=\{qi\}$, then for the purpose of the cyclomatic numbers being equal, it is both necessary and sufficient for all the edge graph $H_q$ vertexes to be simple. 

Theorem has evident corollaries.

\begin{corollary}
If a canonical graph $G_k$ is the adjacency graph for the canonical graph $H_k$ edges, then their cyclomatic numbers are equal: $\nu(G_k )=\nu(H_k)$.
\end{corollary}

\begin {corollary}
If matrix $L$ satisfies to the conditions of the theorem "On the equality of the cyclomatic numbers of both vertex and edge graphs", then it satisfies to the conditions of theorem "On a quasicanonical adjacency matrix".
\end{corollary}

\begin {theorem}
[An extension of the theorem "On the equality of the cyclomatic numbers of both vertex and edge graphs" for the case of the $p$--connected graphs]

Let graoh $G$, which has the $p$ components of the connectedness, be specified by matrix $E$ --- the vertex matrix, which corresponds to the quasicanonical matrix $R_q$ ---- the matrix of graph $H_q$ edges, which also has the $p$ components of the connectedness. For the purpose of equality of both graph cyclomatic numbers, accordingly: $\nu_p (G_p)=\nu_p(H_p)$, it is both necessary and sufficient for all the $r_xy=1$ elements to meet condition (1) of theorem 3.
\end{theorem}

From theorems above it follows that:

\begin {corollary}
If the canonical graph $G_k$ with the $p$ components of the connectedness is adjacency to the edge canonical graph $H_k$ also with the $p$ components of the connectedness, then their cyclomatic numbers are equal.
\end{corollary}

On the basis of the theorems above let us formulate, as evident, a theorem on the canonical binary relation's system.

\begin{theorem}
[On a canonical binary relation's system]
Let set $Q=\{q_i\}$ is given. It has the $p$ components of the connectedness, and the binary relation system $L=Q\times Q$, which has the cyclomatic number $\nu_p(L)$.

Set $Q=\{q_i\}$ and a binary relation system $L=Q\times Q$, adjusted on this set, may be at the same time presented by edge graph $H_k$ with the $p$ components of the connectedness and the cyclomatic number $\nu_p(H_k)=\nu_p(L)$ and its conjugate vertex graph $G_k$ with $p$ components of the connectedness and the same cyclomatic number $\nu_p(G_k)=\nu_p(L)$ if and only if the binary relation system $L=Q\times Q$ has the canonical form, in other words, matrix $\left\|l_{ij}\right\|_1^n=L$  satisfies to theorem 3 conditions. 
\end{theorem}

\begin {theorem}
[On the binary relation's normalization]
Any arbitrary $L=Q\times Q$ system of the binary relation $q_x R q_y$, which is assigned on set $Q=\{q_i\}$, may be brought to the canonical form by the way of the consistent application of $\Delta n$--transformation to that $\left\|l_{ij}\right\|_1^n=L=Q\times Q$ matrix $l_xy=1$ elements, which do not satisfy to theorem 3 conditions.
\end{theorem}

\begin {corollary}
A vertex graph $G$ is the adjacency graph to the edges of some edge graph $H$ if and only if the adjacency matrix of the vertex graph $G$ has either the canonical or the quasicanonical form.
\end{corollary}

\begin{remark}
The consequence does not coincide with the mentioned above Krausz condition \cite{Krausz} since Krausz condition regards to the undirected graphs, and the corollary 4 --- only to the directed graphs.
\end{remark}

\begin {corollary}
If in graph $H(V,Q)$, which has either the canonical or the quasicanonical form, we can pick out such Euler partial graph $H_e(V_e,Q_e)$ that for all of its vertexes $s_h(H_e )=2$, and this graph also has all these edges, that have the one-to-one depentanizer to the initial $G(Q,\Gamma)$ graph's vertexes, then graph $G(Q,\Gamma)$ has Hamilton cycle, and this cycle can be determined unambiguously.
\end{corollary}

\begin {corollary}
A number of Hamilton cycles in graph $G$ is equal to the number of Euler partial graphs in either graph $H_q$ or graph $H_k$, obligatory containing also all the edges that are the one-to-one depentanizer to the $G$ graph's vertexes.
\end{corollary}

\begin {corollary}
If either graph $H_q$ or graph $H_k$ has a family of the noncrossing by the edges the circuits (paths) that cover all those edges, with which graph $G$ vertexes have the one-to-one depentanizer, then to this family of the circuits (paths) in graph $G$ corresponds a family of the non-crossing similar either circuits or chains, which cover all vertexes of graph $G(Q,\Gamma)$. 
\end{corollary}

In the figures below you can see the statements of some proved theorems, which will be necessary in sequel. Corollary 6 appears to be very interesting, because it gives us the conditions of the Hamilton cycles' unambiguous presence in the vertex graph. 

\begin{figure}[htb]
	\includegraphics[width=0.95\textwidth]{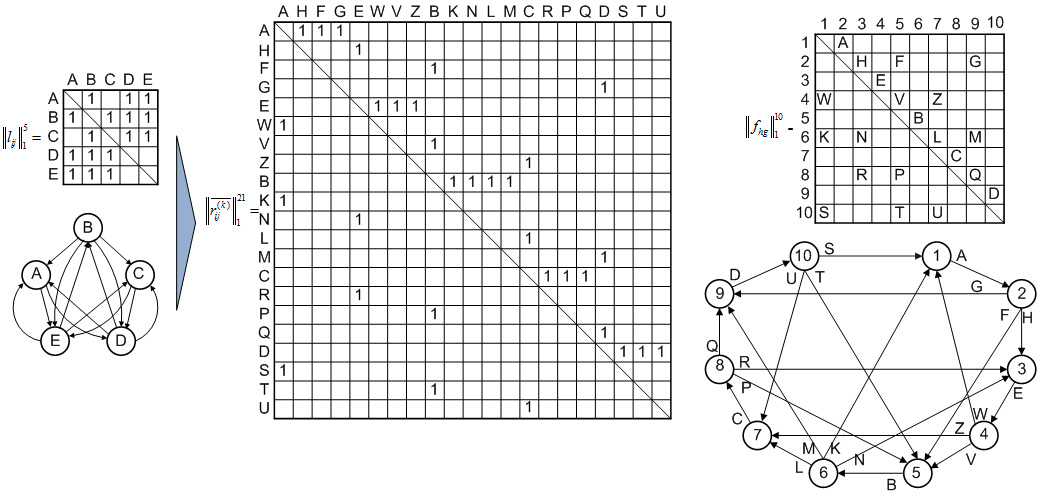}
	\caption{}
	\label{Fig:5}
\end{figure}

\begin{figure}[htb]
	\includegraphics[width=0.85\textwidth]{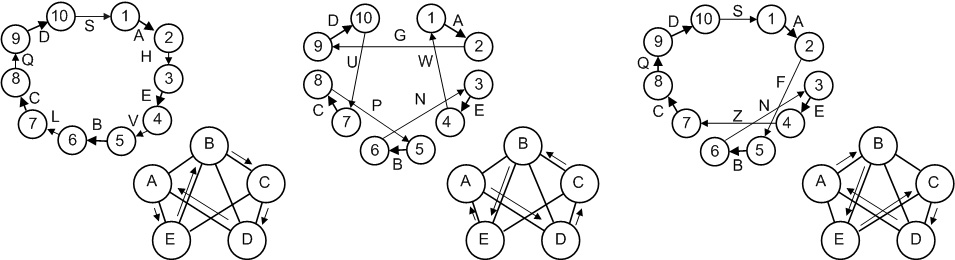}
	\caption{}
	\label{Fig:6}
\end{figure}

As a result of the research of the direct path matrix's properties were proved both necessary and sufficient conditions of both the existence and the uniqueness of the direct path matrix's duality. 

The proved theorems allow solving a problem of the transformation of the arbitrary (both directed and undirected) vertex graphs into the edge graphs. The proved theorems allow formulating the following foundations of the graph's isomorphism:

\begin{enumerate}
\item
The principle of the confluent duality:
\begin{enumerate}
\item
The numbers of the elements of the initial both $Q_0$ and $Q$ sets are equal, that is: $n_(Q_0)=n_Q$; 
\item
The binary relation's systems on both $Q_0$ and $Q$ sets are equal, that is $L_0=L$;
\item
The cyclomatic numbers of both the initial vertex $G$ graph and the edge $H_q$ graph may be different, that are: $\nu(G_q)\geq (H_q)$. 
\end {enumerate}
\item
The principle of the strict duality:
\begin{enumerate}
\item
$n_{Q_0}=n_Q$;
\item
$L_0=L$;
\item
$\nu(G_q)=\nu(H_q)$.
\end{enumerate}
\item
The principle of the normalization: every arbitrary system $L\subset Q\times Q$ of binary relation form $q_i R q_j$, adjusted on set $Q=\{q_i\}$, may be brought to either the semi-canonical or the canonical form with the help of the single structural method of the $\Delta n$--transformation (the operation of the semi-normalization or normalization).
\item
The principle of the reduction: if matrix $L$, corresponding to the binary relation system $L\subset Q\times Q$, adjusted on set $Q=\{q_i\}$, contains such pairs of elements $l_x =1$ and $l_y=1$, which satisfy to the theorem 3, then system $L\subset Q\times Q$ is not obligatory the forming, but always may be converted to the appearance of the forming binary relation's system with the help of the ($-\Delta n)$--transformation (the operation of the reduction).  
\end {enumerate}

As a result of the research of the direct path matrix's properties were proved both necessary and sufficient conditions of both the existence and the uniqueness of the direct path matrix's duality. 

The proved theorems allow solving a problem of the transformation of the arbitrary (both directed and undirected) vertex graphs into the edge graphs.

The proved theorems allow formulating the following foundations of the graph's isomorphism.

As we see the duality between the graphs turns out to be very specific. In general we can see that it is connected with the cyclomatic number and its alteration. Later we'll see how it happens.

\section{A converting of the directed graphs}

Now we will investigate, even briefly, the operation of constructing matrix $F$ (the adjacency matrix of the vertexes of the edge graph $H$) with the help of matrix $R$ (the adjacency matrix of the edges). A lot of what will be said below is presented in \cite{Malinina3} with the proofs of theorems.  

\begin{figure}[htb]
	\includegraphics[width=0.7\textwidth]{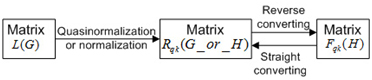}
	\caption{}
	\label{Fig:7}
\end{figure}

This operation is very interesting because it can enlighten the problem of complexity, which is bind with the cyclomatic number. A transfer from matrix $L$ to matrix $R$ may be presented by the scheme in fig. \ref{Fig:7}.

Let us agree to denote the operation of constructing the vertex graph by the given edge graph as the straight converting, and the operation of constructing the edge graph by the given vertex graph --- the reverse converting. Both operations have the following evident features:

\begin{enumerate}
\item
Every directed graph $H$ may be subjected to the consecutive straight converting any number of times.
\item
The directed graph $G$ may be subjected to the consecutive reverse converting if and only if the adjacency matrix of its vertexes, being considered as matrix $R$, has either the quasicanonical or the canonical form.
\item
If grap $H$ is a result of the single straight converting of some graph $P$, then it may be subjected to the reverse converting at least once. If graph $H_j$ is a result of the single straight converting of graph $H_1$ for $(j-1)$ times, then it may be subjected to the operation of the reverse converting at least the $j$ times.
\end{enumerate}

Matrix $F$ must meet the definite requirements; it follows from its definition and from the approach to its receiving. If matrix $F$ is received from the quasicanonical matrix $R_q$, then it is called the quasicanonical, is denoted as $F_q$ and must meet the following requirements:

\begin{itemize}
\item
All the matrix elements, situated on the main diagonal, must be equal to zero;
\item
There could not be more than one empty column (corresponding to the initial vertex) in matrix $F$;
\item
If there is an empty column in matrix $F$, then in the corresponding line there must be exactly one non-zero element;
\item
There could not be more than one empty line (corresponding to the final vertex) in matrix $F$;
\item
If there is an empty line in matrix $F$, then in the corresponding column there must be exactly one non-zero element.
\end{itemize}

If matrix $F$ is received from the canonical matrix $R_k$, then it is called the canonical, denoted as $F_k$ and must, in addition to the listed ones, meet the following requirements:

\begin{itemize}
\item
If there are more than one non-zero element in matrix $F$ line $h$, then there must be exactly one non-zero element in column $g=h$;
\item
If there are more than one non-zero element in matrix $F$ column $g$, then there must be exactly one non-zero element in line $h=g$.
\end{itemize}

Against the type of the converting (either the quasicanonical or the canonical one) we can say that matrix $F$ must meet the following requirements: the first list –-- is for the quasicanonical type of converting. The list of properties below is added if we deal with the canonical type of converting. These properties are obvious (it is a matter of constructing of matrix $F$).

We would not take into consideration graphs with loops, graphs with more than either one entrance or one exit. It is not expected in practical applications. In addition they may be reduced to this form.

\begin{figure}[htb]
	\includegraphics[width=0.35\textwidth]{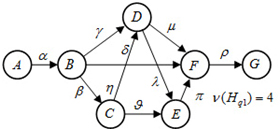}
	\caption{}
	\label{Fig:8}
\end{figure}

Let us take up the example of the straight converting of the arbitrary directed graph. The edge graph $H_{q_1}$, presented in fig. \ref{Fig:8}, is both given and accepted for the initial graph.

We will sequentially transform the edge graph into the vertex one, gathering in the new vertexes first the twains of the vertexes of the initial graph, then the triples, quadruples and so on.

\begin{figure}[htb]
	\includegraphics[width=0.45\textwidth]{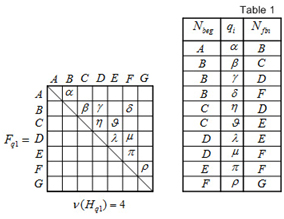}
	\caption{}
	\label{Fig:9}
\end{figure}

A corresponding matrix $F_{q_1}$ is presented in fig. \ref{Fig:9}. Table 1 is arranged according to matrix $F_{q_1}$. An arrangement is evident from matrix $F_{q_1}$ and table 1, and does not need any explanations.

\begin{figure}[htb]
	\includegraphics[width=0.55\textwidth]{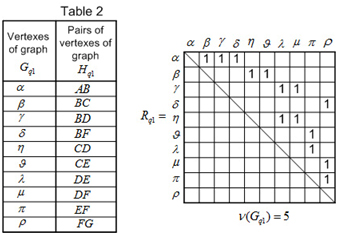}
	\caption{}
	\label{Fig:10}
\end{figure}

Table 2 (fig. \ref{Fig:10}) is arranged accordingly to table 1(fig. \ref{Fig:9}). A difference between table 1 and table 2 is in the fact that both the first and the second columns of table 1 are joined into the second column of table 2. Thus in table 2 an ordered pair of the matrix's rows corresponds to every element $f_{hg}=i$ of matrix $F$. Matrix $R_{q_1}$ is arranged the way you can see below (fig. \ref{Fig:10}).

$$
r_{ij}=
\left\{
        \begin{array}{l}
$
       1,\,if the second element of the ordered pair\, $(N_{beg},N_{fin})$\,
$
\\ 
$
       from the\, $i$\,--line of the table 2 is congruent to the first 
$
\\
$
        element of the ordered pair\, $(N_{beg},N_{fin})$\, from the\,
$
\\
$   
        $j$\,--line of the table 2 
$
\\
        0, \, \,otherwise\,. $\
$
        \end{array}
\right.
$$

\begin{figure}[htb]
	\includegraphics[width=0.65\textwidth]{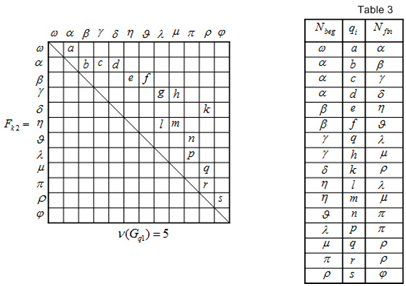}
	\caption{}
	\label{Fig:11}
\end{figure}

Later matrix $R_{q_1}$ is transformed into the operational matrix $F_{k_2}$: elements $r_{ij}=1$ in matrix $R_{q_1}$ are replaced by the arbitrary symbols (for example, $a,b,c...$ and so on) and are added, if it is necessary, one both initial and final row (the initial and the final elements of the basic set of the new graph $H_{k_2}$). Matrix $F_{k_2}$ for our example is presented in fig.\ref{Fig:11}. 

\begin{figure}[htb]
	\includegraphics[width=0.7\textwidth]{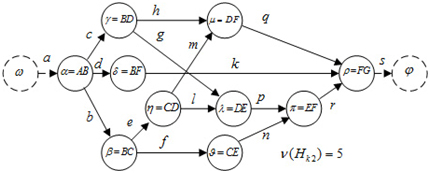}
	\caption{}
	\label{Fig:12}
\end{figure}

A matrix is enlarged with the initial row $\omega$ and the final row $\varphi$. Accordingly both $f_{\omega\alpha}=a$ and $f_{\rho\varphi}=s$ elements are added to the matrix. Graph $H_{k_2}$ is constructed by matrix $F_{k_2}$ (fig. \ref{Fig:12}). A graph appears to be the canonical one, because it has no complicated vertexes. Its cyclomatic number $\nu(H_{k_2})$ is equal to 5, while the cyclomatic number $\nu(H_{q_1})$ of graph $H_{q_1}$ is equal to 4.

\begin{figure}[htb]
	\includegraphics[width=0.85\textwidth]{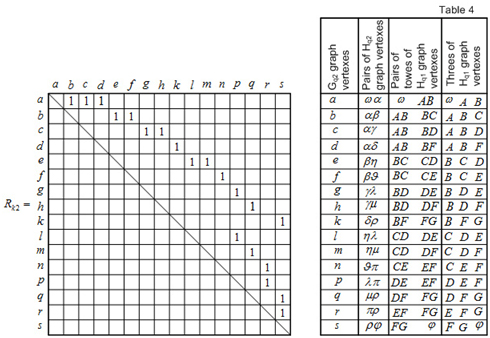}
	\caption{}
	\label{Fig:13}
\end{figure}

Let's execute the converting operation once more. For this we'll arrange table 3 (fig. \ref{Fig:11}) by matrix $F_{k_2}$ similarly like we arranged table 1. The first two columns of table 4 are arranged by table 3 (similarly to table 2). Using table 2, let's decode the ordered pairs of matrix $F_{k_2}$ rows, which are written in the second column of table 4 (fig. \ref{Fig:13}). 

For example, pair $\alpha\beta$ (the second line, the second column of table 4) includes both $\alpha=AB$ and $\beta=BC$ elements. Therefore, in the second line of the third column of table 4 the pairs of deuces of both $AB$ and $BC$ elements are written. All the third column of table 4 is filled up in such a way. 

Let's notice that in all lines of the third column of table 4 the second elements of the first deuces are congruent to the first elements of the second deuces. 

\begin{figure}[htb]
	\includegraphics[width=0.55\textwidth]{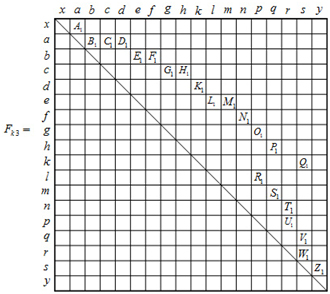}
	\caption{}
	\label{Fig:14}
\end{figure}

Therefore, every ordered pair of the deuces corresponds to the ordered triple of graph $H_{q_1}$ vertexes, excluding both the first and the final ordered triples, each including two vertexes of the initial graph and one vertex, added to graph $H_{k_2}$. For example, the pair of both deuces: $AB$ and $BC$ corresponds to $ABC$ triple. The ordered triples are written in the fourth column of table 4. Matrix $R_{k_2}$ is arranged (fig. \ref{Fig:13}) according to this column of table 4. Then matrix $R_{k_2}$ is transformed into matrix $F_{k_3}$ (fig. \ref{Fig:14}) similarly to matrix $F_{k_2}$ was arranged. 

\begin{figure}[htb]
	\includegraphics[width=0.8\textwidth]{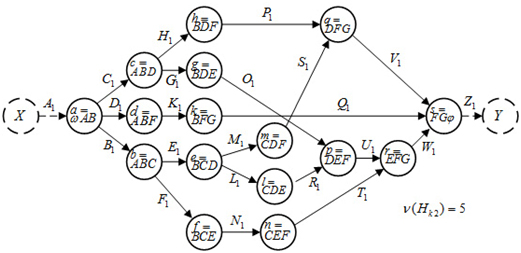}
	\caption{}
	\label{Fig:15}
\end{figure}

Then graph $H_{k_3}$ is constructed by this matrix. It appears to be the canonical graph. Its cyclomatic number is equal to 5. All the vertexes of $H_{k_3}$ graph (excluding both added vertexes $x$ and $y$) correspond to the ordered pairs of graph $H_{k_2}$ vertexes. Every ordered triple of graph $H_{q_1}$ vertexes correspond to some graph $H_{k_3}$ vertex (fig. \ref{Fig:15}).

\begin{figure}[htb]
	\includegraphics[width=0.55\textwidth]{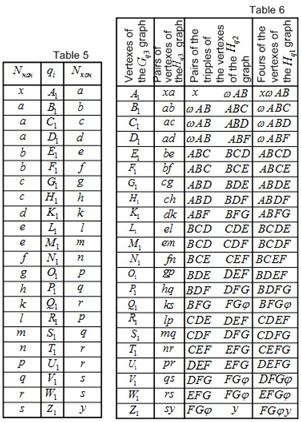}
	\caption{}
	\label{Fig:16}
\end{figure}

Let us execute the converting operation once more. We'll arrange table 5 (fig. \ref{Fig:16}) by matrix $F_{q_3}$ similarly like we arranged tables both 1 and 3. The first two columns of table 6 are arranged by table 5 (similarly to tables both 2 and 4). Using these tables, let's decode the ordered pairs of matrix $F_{q_3}$ rows, which are written in the second column of table 6. For example, pair $ab$ (the second line, the second column of table 6) includes both $a=\omega AB$ and $b=ABC$ elements. Therefore, in the second line of the third column of table 6 by this time the pairs of triples both $\omega AB$ and $ABC$ elements are written. All the third column of table 6 (fig. \ref{Fig:16}) is filled up in such a way.

Let's note that in all the lines of the third column of table 6 two last elements of the first triples are congruent to the first two elements of the second triples. 

Therefore, every ordered pair of the triples corresponds to the ordered four of graph $H_{q_1}$ vertexes, excluding both the first and the last ordered fours, each including two vertexes of the initial graph and two additional. One vertex was added to graph $H_{q_2}$, and the second was added to graph $H_{q_3}$. For example, the pair of both $\omega AB$ and $ABC$ triples correspond to the $\omega ABC$ four.

\begin{figure}[htb]
	\includegraphics[width=0.75\textwidth]{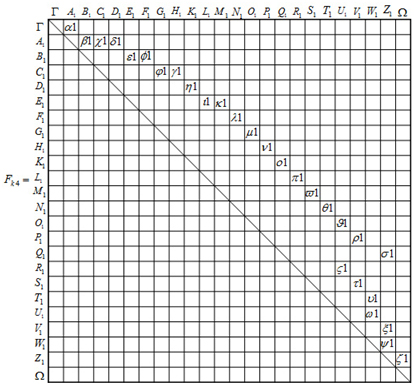}
	\caption{}
	\label{Fig:18}
\end{figure}

The ordered fours are written in the fourth column of table 6. Matrix $R_{k_3}$ is arranged according to table 6. The matrix is transformed into matrix $F_{k_4}$ (fig. \ref{Fig:18}) similarly to both $F_{k_2}$ and $F_{k_3}$ matrixes were arranged. 

\begin{figure}[htb]
	\includegraphics[width=0.55\textwidth]{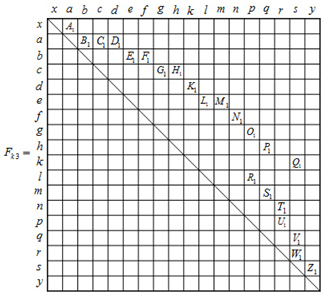}
	\caption{}
	\label{Fig:17}
\end{figure}

\begin{figure}[htb]
	\includegraphics[width=0.9\textwidth]{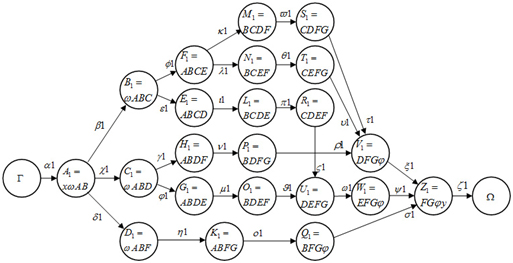}
	\caption{}
	\label{Fig:19}
\end{figure}

Then graph $H_{k_4}$ is constructed by this matrix. It appears to be the canonical graph. Its cyclomatic number is equal to 5. All the vertexes of graph $H_{k_4}$ (excluding the added both $\Gamma$ and $\Omega$ vertexes) correspond to the ordered pairs of graph $H_{k_3}$ vertexes. Every ordered pair of graph $H_{k_1}$ vertexes corresponds to some graph $H_{k_4}$ vertex (fig. \ref{Fig:19}).

So, the straight converting allows constructing a consecutive row of the graphs, which vertexes correspond to the increasingly long orders of the initial graph vertexes. In other words, graph $H_j$, received as a result of the consecutive straight converting of graph $H_1$, being done $(j-1)$ number of times (where $n=j$), possesses the following feature: any graph $H_j$ vertex corresponds uniquely to every ordered $n$ of graph $H_1$ vertexes.

The graphs with the loops we not taken into consideration, graphs with more than either one entrance or one exit were also not taken into consideration. It is not expected in practical applications. In addition they may be reduced to this form.

We sequentially transform the edge graph into the vertex one, gathering in the new vertexes first the twains of the vertexes of the initial graph, then the triples, quadruples and so on. 

Now every ordered quadruple of the initial graph $H_{q_1}$ vertexes correspond to some intermediate graph $H_{k_4}$ vertex.

An examination of such consecutive converting process allows making the following evident conclusions:

\begin {enumerate}
\item
The received $m's$ of the graph $H_1$ vertexes corresponds to all the direct paths of the length $(m-1)$, which are available in graph $H_1$.
\item
Among the received paths there will be all the contours of not more than length $(m-1)$.
\item
While accomplishing the converting $m$ times, we wil get all possible ordered $m's$, among which there will be all graph $H_1$ contours of not more than the $m$ length, including Hamilton circuits.
\end{enumerate}

Thus, the straight converting allows constructing all possible paths and contours of the given length in the initial graph by a rather simple approach. 

But this approach in cases, when the total number of both the paths and the contours in the initial graph is too large and the number of the intermediate graph vertexes at the consecutive converting increases quickly, turns out to be inconvenient for searching either all or the part of the paths of the given form.

Let us see how such increasing of the number of the vertexes of the intermediate graph is connected with the cyclomatic number and the type of the converting. 

\section{The interconnection of the cyclomatic number, the type of the converting and the increasing of the number of the vertexes}

The dependence of the increasing of the number of the vertexes on some $j$--step of the converting was examined and the following theorems were proved.

\begin {theorem}
If graph $H_1$ in the process of its consecutive straight converting generates only the canonical graphs $H_{kj}$ at $(j=1,2,3...M)$, then the numbers of the vertexes of these graphs, received step by step, are determined by the linear dependence from the converting operation's $(j-1)$--number, that is: $n_j=n_1+ \Delta n*(j-1)$ \cite {Malinina3}.
\end{theorem}

\begin {theorem}
If graph $H_1$ in the process of its consecutive straight converting generates both the canonical and the quasicanonical graphs or only the quasicanonical graphs, then the numbers of the vertexes of these graphs, received step by step, are determined by the following expression:

$n_j=n_1+\sum_{\xi=1}^{\xi=(j-1)}\Delta n_{\xi}$ 

Where: $\Delta n_{\xi}=\Delta n_{(\xi-1)}+\Delta \nu(H_{(\xi-1)}$

Where: $\xi=1,2,...(j-1)$, $j=1,2,3,...M$
\end{theorem}

From the formulations of theorems above it follows that all the directed graphs in general may be divided into two classes: 
\begin {enumerate}
\item
The graphs, for which the cyclomatic number always appears to be the invariant of the straight converting on the arbitrary given above converting step.
\item 
All the other directed graphs or graphs, for which the cyclomatic number on either the part or on all the steps of the converting, do not appear to be the invariant of the converting.
\end {enumerate}
 
It is the main conclusion, but a researcher is always a curious person.
 
So, let's proceed to reveal both the \textit{necessary} and the \textit{sufficient} characteristics of these classes of the directed graphs, depending on the concept of the path in the directed graph, which was first suggested by Berge. Simply speaking we want to know what occur with the most part of the directed graphs.

For this purpose theorem below was proved. It declares that the cyclomatic number appears to be the invariant of the converting if the graph is canonical, and all the paths are the holonomic ones. It means that the sum of the vertexes' degrees on the path under study is either the nondecreasing or nonincreasing one.

\begin {theorem}
A cyclomatic number appears to be the invariant of the straight converting of graphs $H_j$ at any $j=1,2,3,...M$ value if and only if graph $H_1$ is canonical and all the paths in graph $H_1$ are the holonomic paths.
\end {theorem}

In other words, the consecutive straight converting of the initial canonical graph $H_1$ with the holonomic paths generates graphs $H_j$ with the same cyclomatic number as the initial graph has. The proof of the theorem is presented \cite {Malinina3}.

Theorem has the evident corollary.

\begin {corollary}
A highest possible $j_{max}$ of the converting steps at $\nu(H_{kj})=Const(j)$, which graph $H_1$ allows, is equal to the length of the shortest of all, belonging to graph $H_1$, the intervals of type $l_{31}$, in other words, it is equal to $min_{\{\xi_1\}}l_{31}$, where all $\xi_1 \in H_1$.
\end{corollary}

\begin{theorem}
Graph $H_{k_1}$ belongs to class $H^{(1)}$ if and only if it does not contain either the contours or the intervals of type $l_{31}$.
\end{theorem}

\begin{theorem}
Graph $H_1$ belongs to class $H^{(2)}$ if and only if it contains the intervals of type $l_{31}$, but does not contain the contours.
\end{theorem}

\begin{theorem}
So as graph $H_1$ was progressive-heteronomous, it is necessary and sufficient for it to contain at least one contour.
\end{theorem}

The last three theorems give us the necessary and sufficient conditions for the graphs to be divided into three classes:
\begin {enumerate}
\item
The first class --- $H^{(1)}$ --- the graphs, which permit the consecutive invariant converting any number of times; 
\item
The second class --- $H^{(2)}$ --- the graphs, for which the cyclomatic number, beginning from some $(j_{min}\geq 0)$ step of the converting, becomes not be the invariant of the converting. But for such graphs we can indicate another, may be very big, but the finite $(j_{kp})$ number of the converting steps, after the accomplishment of which, the cyclomatic number also begins and then continues to be the invariant of the converting, no matter how many times the next converting is accomplished;
\item
The third class --- $H^{(3)}$ --- the graphs, for which, after already accomplished the arbitrarily large number of $j=M$ steps of the preliminary converting, we can indicate  some number of the converting step ($j>M$ steps of the converting), at the accomplishment of which the increasing of the $H_{(j+1)}$ graph cyclomatic number will again occur.
\end{enumerate}
 
\section {Conclusions}

\begin{enumerate}
\item 
If the adjacency vertex matrix of the arbitrary directed $H_1$ graph meets the requirments, which are produced to matrix $F$ --- the adjacency matrix of the vertexes of the edge graph, then graph $H_1$ can be subjected to the operation of the consecutive straight converting any number of times. At that all the graphs are divided as regards to the operation of the consecutive straight converting into three classes:
\begin {itemize}
\item
 	A homonomic
\item
 	A bounded-heteronomous
\item
 	A progressive-heteronomous
\end {itemize}

\item
For the homonomic graphs the cyclomatic number is a regular invariant of the converting independently from the number of the consecutive converting steps. Owing to this fact, the number of the graph $H_j$ vertexes, received from the initial graph $H_1$ as a result of its consecutive converting, has the linear dependence from the number of the steps.
\item
The bounded-heteronomous graphs have the heteronomous bound according to the number of the steps of the consecutive converting. Before the achievement of this bound, the cyclomatic number does not appear to be a regular invariant of converting, but on reaching this bound, as a result of the regular converting step, the homonomic graph is generated, and the cyclomatic number becomes the regular invariant of converting independently from the number of the steps of the following consecutive converting.
\item
The progressive-heteronomous graphs do not have the heteronomous bound according to the number of the steps of the consecutive converting. As a result, the cyclomatic number of the progressive-heteronomous graph never becomes the regular invariant of converting even if the number of the steps will be enormously large. 
\end{enumerate}

\begin{figure}
	\centering
		\includegraphics[width=0.35\textwidth]{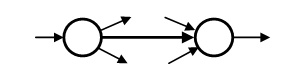}
	\caption{}
	\label{Fig:21}
\end{figure}

Simply speaking, the results may be presented in such words: graph is holonomic and the cyclomatic number appears to be the regular invariant of converting if it has no both the contours and the intervals of such type (fig.\ref{Fig:21}).

Such breakdown of the graphs into three classes and the existence of such a class of the graphs, in which the cyclomatic number appears not to be the invariant of the converting unambiguously shows us why all the attempts to prove $P=NP$ or $P\neq NP$ turn out to be so difficult and ineffective. Thus, concerning the cyclomatic number, we can say that the graphs are divided into two non-overlapping sets: the first and the second classes belong to one set and the third class belongs to another one.

Summarizing all that was said, it can be concluded that such dividing of the graphs into three classes and the behavior of the complicated vertexes at the converting  (they turn into the independent cycles) gives us the infallible fact that it is impossible to prove that $P=NP$. 

So, it may be suggested that $P\neq NP$. 

The breakdown of the graphs in the non-intersecting subsets eventually works as the foundation for such a conclusion.

Above all I can say that the examination of the concrete mass problems, being made according to the proved theorems, will bring to light their concrete peculiarities and I think will help to show how we will be able to manage them. 

In Four Color problem, for example, the problem lies in the vertexes' degrees. In the near future I hope to publish the investigations on the Four Color Hypothesis, which will show us all the difficulties of its proof.

I clearly understand that the problem of $P=NP$ or $P\neq NP$ is too complicated, and you certainly must have a lot of time to suggest whether I'm right or not. 

Also I want to add that a system theory and a theory of the design run into this problem as a physical basis. Topology, logic, combinatory, theory of algorithms, computer science, set theory and etc. serve as auxiliary sciences. But the cardinal science, which I think will give us an answer, will be the graph theory.

Besides I can say that on the base of the proved theorems (\cite {Malinin}, \cite {Malinina}) the algorithms were constructed and it was proved that they are polynomial. Working programs also have been constructed. 

The Chirch-Turing thesis tells us that only normal algorithms may serve as an input data for the computing machines. The computing mashines are constructed in such a mode that only normal algorithms may serve as an input data for them. Otherwise it will not work. It became our severe reality. Computers can do a lot of things, but not everything. 

First it became clearly visible when the algorithm for the Hamilton paths' searching was constructed. In a general case it was not possible to find the Hamilton path effectively, but it turned out that a number of variants for the enumeration were greatly reduced after the operation of the normalization. We were not planning to prove $P=NP?$ (Cook's problem), the result was really un-expected one. To the last moment there was a hope to concur the $NP$--completeness, but unfortunately it came so that it cannot be concurred.

The work was begun and was almost finished by my father, but it was devoted to nets. After his death I was to finish it and I have found the problem in graph theory to which these theorems fit unambiguously. It is the Problem of the Isomorphism between the graphs. And suddenly I understood that these theorems can explain the foundations of the $NP$--completeness, but to some degree. The peculiarities of different mass prob-lems may enlighten their behavior concerning the problem. There may be found much more particular cases and etc.

But nevertheless $P\neq NP$.

\section{Acknowledgement}
The report was made long ago, may be now I see the problem more distinctly, but nevertheless it is a deeply complicated problem, so I apologize, if I cannot talk on it in a simple way.
I also will be very grateful to those readers, who will find and send me a word about the uncovered misprints and withholding facts in order to improve the text.


\begin{thebibliography}{99}

\bibitem {Malinina1}
\textsc {Malinina, N.}:\ \textit {On a principle impossobilyty to prove $P=NP$} International Mathematical Congress, Hyderabad, India, 414-415, (2010) 
   
\bibitem {Malinin}
\textsc {Malinin, L.,Malinina, N.}:\ \textit {Graph isomorphism in theorems and algorithms}, Librocom Publishing Company, Moscow, 248pp., (2009)

\bibitem {Malinina}
\textsc {Malinin, L.,Malinina, N.}:\ \textit{On the soluton of the Graph Isomorphism Problem, Part I}, ArXiv (Cornell University Library), http://arxiv.org/abs/1007.1059, 46p., (2010)                        
\bibitem{Cook}
\textsc{Cook, S. A.}:\ \textit{The complexity of theorem-proving procedures}, Proc. 3rd Ann. ACM Symp. on Theory of Computing, Association for Computing Machinery, New York, 151-158, (1971)

\bibitem {Gary and John}
\textsc {Gary, M.R.,Johnson, D. S.}:\ \textit{Computers and Intractability}, Bell Telephone Laboratories, Inc., (1979)

\bibitem {Turing} 
\textsc {Turing, A.}:\ \textit{On computable numbers, with an application to the Entscheidungsproblem}, Proc. London Math. Soc. Ser., \textbf{2 42}, 230-265 and \textbf{43}, 544-546, (1936)

\bibitem {Markov}
\textsc{Markov, A. A., Nagorniy, N. M.}:\ \textit{Theory of Algorithms}, Phazis, Moscow,  (1996), (Russ.lang.)
                                              
\bibitem {Berge}
\textsc {Berge, C.}:\ \textit {Theore des graphes et ses applications}, DUNOD, Paris, (1958)

\bibitem {Krausz}
\textsc {Krausz, J.}:\ \textit {Demonstration nouvelle d'une theoreme de Whitney sur les resaux} Math. Fiz. Lapok, \textbf{50}, 75-85,(1943)


\bibitem {Malinina3}
\textsc {Malinina, N.}:\ \textit {A converting of the directed graphs}, ArXiv (Cornell University Library), http://arxiv.org/abs/1210.6088, 29p., (2012)

\end{thebibliography}
\end{document}